\documentclass[10pt]{article}
\usepackage{amsfonts}
\usepackage{epsfig,cite}
\begin{document}
\begin{center}
%%%%%%%%%%%%%%%%%%%%%%%%%%%%%%%%%%%%%%%%%%%%%%%%%%%%%%%%%%%%%%%%%%%%%%%%%%%
{\Large\bf Two-Photon Exchange Contribution to Proton Form Factors\\%
\vspace{0.3cm} %
 in Time-Like region}\\
%%%%%%%%%%%%%%%%%%%%%%%%%%%%%%%%%%%%%%%%%%%%%%%%%%%%%%%%%%%%%%%%%%%%%%%%%%%
\vspace*{1cm}
%%%%%%%%%%%%%%%%%%%%%%%%%%%%%%%%%%%%%%%%%%%%%%%%%%%%%%%%%%%%%%%%%%%%%%%%%%%
D. Y. Chen $^1 \footnote{E-mail: chendy@mail.ihep.ac.cn}$, H. Q.
Zhou $^2$ and Y. B. Dong$^1$\\
\vspace{0.3cm} %
{\ $^1$Institute of High Energy Physics \\
 The Chinese Academy of Science,\ Beijing,\ 100049,\ P. R. China}\\
{\ $^2$ Department of Physics, Southeast University,
 Nanjing,\ 211189,\ P. R. China}
%%%%%%%%%%%%%%%%%%%%%%%%%%%%%%%%%%%%%%%%%%%%%%%%%%%%%%%%%%%%%%%%%%%%%%%%%%%
\vspace*{1cm}
\end{center}
%%%%%%%%%%%%%%%%%%%%%%%%%%%%%%%%%%%%%%%%%%%%%%%%%%%%%%%%%%%%%%%%%%%%%%%%%%%
\begin{abstract}
We estimate two-photon exchange contribution to the process $e^+ +
e^- \rightarrow p + \bar{p}$. The two-photon exchange corrections to
double spin polarization observables and form factors in the
time-like region are calculated. The corrections are found to be
small in magnitude, but with a strong angular dependence at fixed
momentum transfer. These two features are the same as those in the
space-like region. In the view of experiment, the double spin
polarization observable $P_z$ deserves to be considered.
\end{abstract}

\section{Introduction}

The electromagnetic form factors in both space-like $(Q^2>0)$ and
time-like $(Q^2<0)$ regions are essential to understand the
intrinsic structure of hadrons. The experimental data of elastic
form factors over several decades, including recent high precision
measurement at Jefferson Lab \cite{PRL-1398,PRL-092301} and
elsewhere \cite{PRD-5491}, have provided considerable insight into
the detail structure of the nucleon. Generally, in Born amplitude
for one photon exchange, the proton current operator is
parameterized in terms of Dirac $(F_1)$ and Pauli $(F_2)$ form
factors,
\begin{eqnarray}
\Gamma_{\mu}=F_1(q^2)\gamma_{\mu}+i\frac{F_2(q^2)}{2 m_N} \sigma_{\mu
\nu} q^{\nu},
\label{ff0}%
\end{eqnarray}
where $q$ is the momentum transfer to the nucleon and $m_N$ is the
nucleon mass. The resulting differential cross section depends on
two kinematic variables, conventionally taken to be $Q^2\equiv -q^2$
(or $\tau$, in order to consistent with the case in the time-like
region, we take $\tau \equiv q^2/4m_N$ other than $\tau \equiv
Q^2/4m_N$ ) and the scattering angle $\theta_e$ (or virtual photon
polarization $\varepsilon \equiv [1+ 2(1-\tau) \tan^2
(\theta_e/2)]^{-1}$). The reduced Born cross section, in terms of
the Sachs electric and magnetic form factors, is
\begin{eqnarray}
\frac{d\sigma}{d\Omega}=C(Q^2,\varepsilon)
\left[G_M^2(Q^2)-\frac{\varepsilon}{\tau} G_E^2(Q^2)\right].
\label{cs0}%
\end{eqnarray}

\par%

The standard method that has been used to determine the electric and
magnetic form factors, particularly those of the proton has been the
Rosenbluth, or longitudinal-transverse(LT), separation method. The
results of the Rosenbluth measurements for the proton form factor
ratio $R=\mu_p G_E/G_M$ have generally been consistent with $R
\approx 1$ for $Q^2\leq 6 GeV^2$
\cite{PRD-5671,PRC-034325,PRC-015206}. The 'Super-Rosenbluth'
experiment at Jefferson Lab \cite{PRL-142301}, with very small
systematic errors were achieved  by detecting the recoiling proton
rather than the electron, is also consistent with the earlier LT
results. It should be mentioned that polarized lepton beams give
another way to access the form factors \cite{SP-588}. In the Born
approximation, the polarization of the recoiling proton along its
motion $(p_l)$ is proportional to $G_M^2(Q^2)$ while the component
perpendicular to the motion $(p_t)$ is proportional to
$G_E(Q^2)G_M(Q^2)$. Then the form factor ratio $R$ can be determined
through a measurement of $p_t/p_l$, with
\begin{eqnarray}
\frac{p_t}{p_l}=-\sqrt{-\frac{2\varepsilon}{\tau(1+\varepsilon)}}
\frac{G_E(Q^2)}{G_M(Q^2)} \  .
\end{eqnarray}
This method has been applied only recently in Jefferson Lab
\cite{PRL-1398}, since it needs high-intensity polarized beams,
large solid-angle spectrometers, and advanced techniques of
polarimetry in $GeV$ range. The measurement about the
electron-to-proton polarized transfer in $\vec{e}^{\ -} + p
\rightarrow e^{-}+ \vec{p}$ shows that the ratio of Sachs form
factors \cite{PR-2256,NC-821} is monotonically decreasing with
increasing of $Q^2$, which strongly contradicts to the scaling ratio
determined by the traditional Rosenbluth separation method
\cite{PR-615}. In order to explain the discrepancy, radiative
corrections, especially the two-photon contribution, have been
involved \cite{PRC-054320, PRL-142303, PRL-142304,
PRL-172503,PRL-122301, PRC-065203, PRC-038202}. In Ref.
\cite{PRL-142304}, only the intermediate proton state considered, it
is found that the two-photon corrections have the proper sign and
magnitude to resolve a lager part of the discrepancy between the two
experimental techniques. Furthermore, Ref. \cite{PRL-172503}
considered the intermediate $\Delta^+$ state as well as the proton.
In Ref.\cite{PRL-122301} a partonic calculation of the two-photon
exchange contribution to the form factors is given. It is concluded
that for $Q^2$ in the range of $2 \sim 3 GeV^2$, the ratio extracted
using LT method including the two-photon corrections agrees well
with the polarization transfer results. Consequently, it shows that
the two-photon exchange corrections can, at least, partly explain
the discrepancy of the two methods of the separation.

\par%

For a stable hadron, in the space-like region the form factors are
real, while its time-like form factors have a phase structure
reflecting the final-state interactions of the outgoing hadrons,
therefore, form factors are complex. So far, there are not many
precise experimental data in this region as in the space-like one.
In the theoretical point of view, it seems unavoidable to check the
two-photon exchange contribution to the nucleon form factors in the
time-like region. Actually, some works have been done. Refs.
\cite{PRC-042202, NPA-120, PLB-197} employed the general arguments
based on crossing symmetry for the processes of  $e^{-} + h
\rightarrow e^{-} +h$ and $e^{+} + e^{-} \rightarrow h + \bar{h}$,
and showed the general expressions for the polarization observables
of  the reaction $\bar{p} + p \rightarrow e^{+} + e^{-}$ in terms of
three independent complex amplitudes and in presence of two-photon
exchange. Ref. \cite{PLB-197} also tried to search some experimental
evidences for the two-photon exchange from the experimental data of
$e^{+} + e^{-} \rightarrow p+\bar{p}+\gamma$. However, a negative
conclusion is obtained due to the level of the present precision. A
total contribution of the radiative corrections to the angular
asymmetry is under $2 \%$, while the asymmetry getting from the
experimental data is always compatible with zero and the typical
error is about $5 \%$. In this reference, the polarization
observables are not discussed.

\par%

Difference with the above work, we calculate the two-photon exchange
correction to the unpolarized differential cross section as well as
the double spin polarization observables. Some qualitative
properties based on the crossing symmetry and C- invariance are
discussed in section $2$. Moreover, the analytical forms of the
unpolarized differential cross section and polarization observables
are presented in section $3$. In section $4$, we will directly
calculate the two-photon exchange contribution to the differential
cross section and polarization observables. In section $5$, some
numerical results and discussions are given.

\par%

\section{Crossing Symmetry and C-invariance}

\begin{figure}[h]
\centering
\mbox{\epsfig{figure=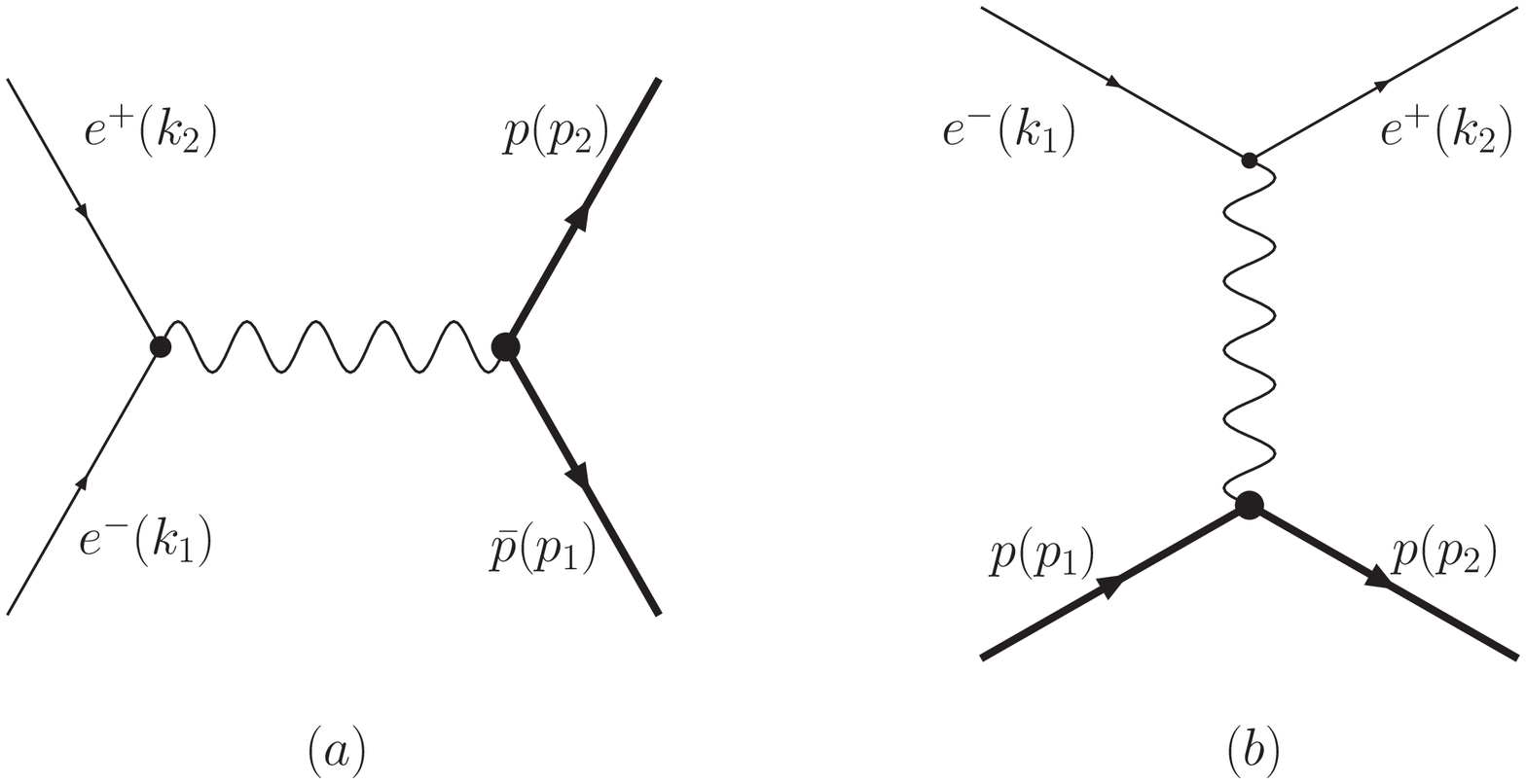,width=120mm,clip=}} %
\renewcommand{\figurename}{Fig.}
\caption{One-photon approximation for the crossed channels. The left
one represents the annihilating  channel of $e^{+} +
e^{-}\rightarrow h + \bar{h}$, and the right one shows the
scattering channel of $e^{-} + h\rightarrow e^{-} + h$.}
\label{Fig-feyntree}%
\end{figure}

In quantum field theory, crossing symmetry is a symmetry that relates to the
$S$-matrix elements. In general, the $S$-matrix for any
process involving a particle with momentum $p$ in the initial state
is equal to the $S$-matrix for an otherwise identical process but
with an anti-particle of momentum $k=-p$ in the final state, that is,
\begin{eqnarray}
\mathcal{M}(\phi(p)+\cdot\cdot\cdot\rightarrow\cdot\cdot\cdot)=
\mathcal{M}(\cdot\cdot\cdot\rightarrow\cdot\cdot\cdot+\bar{\phi}(k)),
\end{eqnarray}
where $\bar{\phi}$ stands for anti-particle and $k = -p$. We notice
that there is no any realistic value of $p$ for which $p$ and $k$
are both physically allowed. So technically we should say that
either amplitude can be obtained from the other by analytic
continuation. The crossing symmetry provides a relation between the
scattering channel $e^{-} + p \rightarrow e^{-} + p$ and the
annihilating  channel $e^{+} + e^{-} \rightarrow p + \bar{p}$. In
the one-photon approximation as shown in Fig. (\ref{Fig-feyntree}),
the crossing symmetry can be expressed by the following relation
\begin{eqnarray}
\overline{|\mathcal{M}(e^{-}p \rightarrow
e^{-}p)|^2}=f(s,t)=\overline{|\mathcal{M}(e^{+} e^{-} \rightarrow p
\bar{p})|^2}.
\label{csym}%
\end{eqnarray}
The line over $\mathcal{M}$ denotes the sum over the polarization of
all particles in the initial and final states. The Mandelstan
variables $s$ and $t$ are defined as follows:
\begin{eqnarray}
s&=&(k_1+p_1)^2=m_N^2+2 E_1 m_N \geq m_N^2,\nonumber\\
t&=&(k_1-k_2)^2=q^2<0,
\end{eqnarray}
for the scattering channel (with  $E_1$ being the energy of the
incoming electron in the Lab frame), and
\begin{eqnarray}
s&=&(k_1-p_1)^2=m_N^2-2 \widetilde{\epsilon}^2+2 \widetilde{\epsilon}
\sqrt{\widetilde{\epsilon}^2-m_N^2} \cos \theta \leq 0,\nonumber\\
t&=&(k_1+k_2)^2=4 \widetilde{\epsilon}^2>4 m_N^2,
\end{eqnarray}
for the annihilating  channel with $\widetilde{\epsilon}$ being the energy of
the initial electron (or final proton) and $\theta$ being the hadron
production angle.

\par%
Considering Lorentz, parity, time-reversal, and helicity
conservation in the limit of $m_e\rightarrow 0$,  the $T-$ matrix
for the elastic scattering of two Dirac particles can be expanded in
terms of three independent Lorentz structures. Then, the proton
current operator through the Lorentz structure \cite{SP-588} is
\begin{eqnarray}
\Gamma_{\mu}=\widetilde{F}_1(s,t)\gamma_{\mu}+i\frac{\widetilde{F}_2(s,t)}{2
m_N} \sigma_{\mu \nu} q^{\nu} +\widetilde{F}_3(s,t) \frac{\gamma \cdot
K P_{\mu}}{m_N^2},
\label{ff1}%
\end{eqnarray}
with
\begin{eqnarray}
P=\frac{1}{2} (p_2+p_1),\ \ \ \ \  K=\frac{1}{2} (k_1+k_2),
\end{eqnarray}
in the scattering channel, and
\begin{eqnarray}
P=\frac{1}{2} (p_2-p_1),\ \ \ \ \  K=\frac{1}{2} (k_1-k_2),
\end{eqnarray}
in the annihilating  channel. Similar to the Sachs form factor, we can
recombine the form factors $\widetilde{F}_{1,2}$ as
\begin{eqnarray}
\widetilde{G}_E(q^2, \cos \theta) &=&\widetilde{F}_{1} (q^2, \cos
\theta) +\tau \widetilde{F}_2(q^2, \cos \theta),\nonumber\\
\widetilde{G}_M(q^2, \cos \theta) &=&\widetilde{F}_{1} (q^2, \cos
\theta) + \widetilde{F}_2(q^2, \cos \theta).
\label{sachs1}%
\end{eqnarray}

\par%

Taking the proton current operator defined in Eq. (\ref{ff1}) which
includes the multi-photon exchange, we can express $f(s,t)$ in Eq.
(\ref{csym}) in the form:
\begin{eqnarray}
f(s,t)=\frac{8 e^4}{(4 m_N^2 -t)t}\Big\{8 |\widetilde{G}_E|^2 m_N^2
\big[m_N^4-2 s m_N^2 + s(s+t)\big]- |\widetilde{G}_M|^2 t\big[2
m_N^4 - 4 m_N(s+t)+2 s^2+t^2 +2 s t\big] \nonumber\\
-m_N^{-2} \big[2 m_N^6-m_N^4(6s+t)+2 m_N^2s(3s+2t)-s(2s^2+3ts+t^2)
\big]Re\big[(4 m_N^2 \widetilde{G}_E-t
\widetilde{G}_M)^{*}\widetilde{F}_3\big]\Big\}.
\end{eqnarray}

\par%

In the one-photon mechanism for $e^{+} + e^{-} \rightarrow p +
\bar{p}$, the conservation of the total angular momentum
$\mathcal{J}$ allows only one value of $\mathcal{J}=1$. This is  due to the
quantum numbers of the photon : $\mathcal{J}^p=1^-, C(1 \gamma)=-1$.
The selection rule combined with  $C$ and $P$ invariances allows
two states for $e^{+} e^{-}$ (and $p \bar{p}$):
\begin{eqnarray}
S=1,\ \ \ \ell=0\ \ \  \mathrm{and}\ \ \ \ S=1,\ \ \ \ \ell=2 \ \ \
\ \mathrm{with}\ \ \ \mathcal{J}^p=1^{-},
\end{eqnarray}
where $S$ is the total spin and $\ell$ is the orbital angular
momentum of the $e^{+} e^{-}$ (or $p \bar{p}$) system. As a result
the $\theta$ dependence of the differential cross section for $e^{+}
+ e^{-} \rightarrow p + \bar{p}$, in the one-photon exchange
mechanism, has the following general form
\begin{eqnarray}
\frac{d\sigma^{1 \gamma}}{d \Omega} = a(t) +b(t) \cos^2 \theta.
\label{sigma0}%
\end{eqnarray}
Similar analysis can be done for the $\cos \theta$ dependence of the
$1 \gamma \otimes 2 \gamma-$ interference contribution to the
differential cross section of this precess. In general, the spin and
parity of the $2 \gamma -$ states are not fixed, but only a positive
$C-$ parity, $C ( 2 \gamma ) = +$, is allowed, then the $\cos
\theta$ dependence of the $1\gamma \otimes 2 \gamma$ interference
contribution to the differential cross section can be predicted on
the basis of its $C-$ odd nature as:
\begin{eqnarray}
\frac{d\sigma^{int}}{d\Omega}= \cos\theta \big[c_0(t)+c_1(t) \cos^2
\theta + c_2(t) \cos^4 \theta + ...\big].
\label{sigmaint}%
\end{eqnarray}

\par%

In the one-photon exchange mechanism, the differential cross section
is angular symmetric. However, after considering the two-photon
exchange, this symmetry is  broken. Define the asymmetry of the
total differential cross section as
\begin{eqnarray}
A_{2 \gamma}(q^2,\theta)= \frac{\displaystyle\frac{d \sigma}{d
\Omega}(q^2, \theta) - \frac{d \sigma}{d \Omega}(q^2,\pi-\theta) }
{\displaystyle\frac{d \sigma}{d \Omega}(q^2, \theta) + \frac{d
\sigma}{d \Omega}(q^2,\pi-\theta) }\ \ ,
\end{eqnarray}
after some algebraic simplification, we have
\begin{eqnarray}
A_{2\gamma}(q^2, \theta)=\frac{d\sigma^{int}}{d \Omega} (q^2,\theta)
~\Big/~ \frac{d\sigma^{1 \gamma}}{d \Omega}(q^2,\theta).
\end{eqnarray}
Then based on the general forms of $d \sigma^{1\gamma}/d\Omega$ and
$d \sigma^{int}/d\Omega$ as shown in Eq. (\ref{sigma0}) and Eq.
(\ref{sigmaint}), One can easily conclude that the angular asymmetry
of the total differential cross section is also an odd function of
$\cos\theta$.

\section{Differential Cross Section and Polarization Observables}
\label{dcspo}%

In order to represent the polarization vector of outgoing anti-proton
in a straight way for the process of $e^{+} + e^{-} \rightarrow p +
\bar{p}$, we define a coordinate frame in center of mass system (CMS)
of the reaction in such a way that the $z$ axis directs along the
three-momentum of the anti-proton and the angle between the incoming
electron and outgoing anti-proton is defined as $\theta$. In such a
frame, according to the approaches used in Refs. \cite{EPJA-331,
NPA-271, NPA-322}, one has
\begin{eqnarray}
\mathcal{M}=\frac{e^2}{q^2} j_{\mu} J^{\mu}
\end{eqnarray}
with leptonic current
\begin{eqnarray}
j_{\mu}=\bar{u} (-k_2) \gamma_{\mu} u(k_1) \nonumber
\end{eqnarray}
and hadronic current
\begin{eqnarray}
J_{\mu}=\bar{u} (p_2)
\Big[\widetilde{F}_1(s,t)\gamma_{\mu}+i\frac{\widetilde{F}_2(s,t)}{2
m_N} \sigma_{\mu \nu} q^{\nu} +\widetilde{F}_3(s,t) \frac{\gamma
\cdot K P_{\mu}}{m_N^2}\Big] u(-p_1)
\label{current}%
\end{eqnarray}

Then the differential cross section of the reaction in the CMS is
\begin{eqnarray}
\frac{d \sigma}{d \Omega} =\frac{\alpha^2\beta}{q^6}L_{\mu \nu}
H^{\mu \nu}, \ \ \ L_{\mu \nu} =j_{\mu} j^{*}_{\nu}, \ \ \ H_{\mu
\nu} = J_{\mu} J^{*}_{\nu},
\label{tensor}%
\end{eqnarray}
$\alpha=e^2/4 \pi$ is the fine structure constant and
$\beta=\sqrt{1-4M^2/q^2}$ is the nucleon velocity in the CMS. In
this work we consider the unpolarized incoming positron and
longitudinally polarized incoming electron with the polarization
four-vector $s$, and in the final state, the anti-proton is
polarized with polarization four-vector $s_1$, then the leptonic and
hadronic vectors can be divided into unpolarized and polarized parts
\begin{eqnarray}
L_{\mu\nu} = L_{\mu \nu}(0) +L_{\mu\nu} (s), \ \ \  H_{\mu \nu}
=H_{\mu \nu}(0) +H_{\mu \nu} (s_1).
\end{eqnarray}
\par%

In the current operator shown in Eq. (\ref{ff1}), the Lorentz
structure functions are not only the function of $q^2$ but also
depend on hadron production angle $\theta$, and they can relate to
the Dirac and Pauli form factors
\begin{eqnarray}
\widetilde{F}_{1,2} (q^2, \cos\theta)=F_{1,2} (q^2) + \Delta F_{1,2}
(q^2, \cos\theta)
\end{eqnarray}
and $\widetilde{G}_{E,M}(q^2, \cos \theta)$ related to the Sachs
form factors
\begin{eqnarray}
\widetilde{G}_{E,M}(q^2, \cos \theta)= G_{E,M}(q^2) + \Delta
G_{E,M}(q^2, \cos \theta) \ .
\label{dsachs}%
\end{eqnarray}
\par%

The unpolarized differential cross section of the process $e^{+} +
e^{-} \rightarrow p + \bar{p}$ is in the form
\begin{eqnarray}
\frac{d\sigma_{un}}{d\Omega}=\frac{\alpha^2 \beta}{4 q^6}L_{\mu
\nu}(0) H^{\mu \nu}(0)=\frac{\alpha^2 \beta}{4 q^2}\  D,
\end{eqnarray}
with the current operator in Eq. (\ref{ff1}) and the definition in
Eq. (\ref{sachs1}), $D$ can be expressed as:
\begin{eqnarray}
D=|\widetilde{G}_M|^2 (1+ \cos^2\theta) +\frac{1}{\tau}
|\widetilde{G}_E|^2 \sin^2\theta - 2 \sqrt{\tau (\tau-1)}
Re[(\widetilde{G}_M-\frac{1}{\tau} \widetilde{G}_E) \widetilde{F}_3
^{*}] \sin^2\theta \cos \theta.
\end{eqnarray}
Notice that in Eq.(\ref{dsachs}), $\Delta G_{E,M}$ and
$\widetilde{F}_3$ caused by the two-photon exchange is in the order
of $\alpha \simeq 1/137$, so that the terms $\Delta G_{E,M} \Delta
G_{E,M}$ and $\Delta G_{E,M}\widetilde{F}_3$ are negligible, then,
\begin{eqnarray}
D&=&|G_M|^2 (1+\cos^2 \theta) + \frac{1}{\tau} |G_E|^2 \sin^2 \theta
+ 2 Re[G_M \Delta G^*_M] (1+ \cos^2 \theta) + \nonumber\\
&&\frac{2}{\tau} Re[G_E \Delta G^*_E] \sin^2 \theta- 2 \sqrt{\tau
(\tau-1)} Re[(G_M-\frac{1}{\tau} G_E) \widetilde{F}_3^{*}]
\sin^2\theta \cos \theta.
\label{dd}%
\end{eqnarray}
\par%

From $C-$ invariance and the above expression of $D$, we have the
general properties of the form factors,
\begin{eqnarray}
\Delta G_{E,M}(q^2, +\cos\theta)&=&-\Delta
G_{E,M}(q^2,-\cos\theta),\nonumber\\ \widetilde{F}_3(q^2, +\cos
\theta) &=& \widetilde{F}_3(q^2, -\cos \theta).
\label{gemf3}%
\end{eqnarray}
which is equivalent to the symmetry relations of the scattering channel
\cite{EPJA-331}.

Generally, the polarization four-vector $S_{\mu}$ of a relativistic
particle with three-momentum $\vec{p}$ and mass $m$, is connected
with the polarization vector, $\vec{\xi}$, by a Lorentz boost:
\begin{eqnarray}
\vec{S} = \vec{\xi} + \frac{\vec{p}\cdot\vec{\xi}\ \vec{p}}{m(E+p)}\
,\ \ \ \ \ S^0 =\frac{\vec{p}\cdot \vec{S}}{m}\ .
\end{eqnarray}
Where $E=\sqrt{m^2+\vec{p}^2}$ is the energy of the particle. In the
CMS defined above, we have the polarization vectors of the
anti-proton
\begin{eqnarray}
\vec{\xi}_x &=&(1,\ 0,\ 0),\ \ \ \  s_{1x}= (0,\ 1,\ 0,\ 0),
\nonumber\\
\vec{\xi}_y &=&(0,\ 1,\ 0),\ \ \ \  s_{1y}= (0,\ 0,\ 1,\ 0),
\nonumber\\
\vec{\xi}_z &=&(0,\ 0,\ 1),\ \ \ \  s_{1z}= (\sqrt{\tau -1},\ 0,\
0,\ \sqrt{\tau}).
\end{eqnarray}

\par%
$P_y$ is a single-spin polarization observable, which relates to one
polarized particle along the $y-$ axis. Since the time-like form
factors are complex, then it appears in the Born approximation in
the process $e^{+} + e^{-} \rightarrow p + \bar{p}$. In this work we
consider the outgoing anti-proton polarized. The general expression
for $P_y$ is
\begin{eqnarray}
P_y=\frac{1}{D q^4} L_{\mu\nu} H_{\mu \nu} (s_{1 y})=\frac{1}{D q^4}
\big[L_{\mu\nu}(0) H_{\mu \nu} (s_{1 y})+L_{\mu\nu}(s) H_{\mu \nu}
(s_{1 y})\big].
\label{py}%
\end{eqnarray}
After some algebraic calculation \cite{arXiv-0704.3375}, we can find
$P_y$ does not depend on the polarization of incoming electron, that
means the second term in Eq. (\ref{py}) has no contribution to
$P_y$. With the proton current operator in Eq. (\ref{ff1}) we have,
\begin{eqnarray}
P_y&=& \frac{2 \sin \theta}{D \sqrt{\tau}} \big[Im[\widetilde{G}_M
\widetilde{G}^{*}_E] \cos \theta - \sqrt{\tau (\tau -1)} (Im[
\widetilde{G}_E \widetilde{F}^{*}_3 ] \sin^2 \theta +Im[
\widetilde{G}_M \widetilde{F}^{*}_3 ] \cos^2 \theta )\big]\nonumber\\
&=& \frac{2 \sin \theta}{D \sqrt{\tau}} \big[Im[G_M G^{*}_E + G_M
\Delta G^{*}_E+\Delta G_M G^{*}_E] \cos \theta -\nonumber\\
&&\sqrt{\tau (\tau -1)} (Im[ G_E \widetilde{F}^{*}_3 ] \sin^2 \theta
+Im[G_M \widetilde{F}^{*}_3 ] \cos^2 \theta ) \big].
\label{polarpy}%
\end{eqnarray}
Similar definitions are employed for the double spin polarization
observables $P_x$ and $P_z$. For $P_x$  and $P_z$, the polarization
of the incoming electron is necessary, and the unpolarized incoming
electron has no contribution, that is $L_{\mu\nu}(0) H_{\mu \nu}
(s_{1x,z})=0$. Since $L_{\mu \nu}(0) H^{\mu \nu} (s_{1x,z}) \propto
a_{\mu}b_{\nu}c_{\rho}d_{\lambda}\epsilon^{\nu\nu\rho\lambda}\equiv
\epsilon^{a b c d}$ and $a,\ b,\ c,\ d$ are four out of $s_{1x,z},\
k_1,\ k_2,\ p_1,\ p_2$, and all of  those four-vectors have zero
$y-$ component, then the contribution of the unpolarized electron
vanishes. The double spin polarization observables with proton
current operator in Eq. (\ref{ff1}) are
\begin{eqnarray}
P_x&=&-\frac{2 \sin \theta}{D \sqrt{\tau}} \Big\{Re[\widetilde{G}_M
\widetilde{G}^{*}_E] + Re[\widetilde{G}_M \widetilde{F}^{*}_3]
\sqrt{\tau (\tau-1)} \cos \theta \Big\}\nonumber\\
&=&-\frac{2 \sin \theta}{D \sqrt{\tau}} \Big\{Re[ G_M G^{*}_E + G_M
\Delta G^{*}_E+\Delta G_M G^{*}_E] + Re[G_M \widetilde{F}^{*}_3
]\sqrt{\tau (\tau-1)} \cos \theta \Big\},\nonumber\\
P_z&=&\frac{2}{D}\Big\{|\widetilde{G}_M|^2 \cos \theta
-Re[\widetilde{G}_M \widetilde{F}_3^{*} ] \sqrt{\tau (\tau-1)}\sin^2
\theta \Big\}\nonumber\\
&=&\frac{2}{D}\Big\{|G_M|^2 \cos \theta + 2 Re[G_M \Delta G_M] \cos
\theta -Re[G_M \widetilde{F}_3^{*} ] \sqrt{\tau (\tau-1)}\sin^2
\theta \Big\}.
\label{polar}%
\end{eqnarray}
In Eqs. (\ref{polarpy},\ref{polar}), if we set $\Delta G_{E,M}=0$
and $\widetilde{F}_3=0$, the polarization observables reduces to the
results in the one-photon approximation. Considering the two-photon
exchange contribution to the double spin polarization observables,
we define $\delta(P_{x,z})$ as the ratio between the contributions
of $1 \gamma \otimes 2 \gamma$ interference terms and the results in
the one-photon mechanism, that is ,
\begin{eqnarray}
\delta(P_{x,z}) =P^{int}_{x,z}/P^{1 \gamma}_{x,z},\nonumber
\end{eqnarray}
with Eq. (\ref{polar}) we have,
\begin{eqnarray}
\delta(P_x) &=& \frac{Re[G_M \Delta G^{*}_E + G_E \Delta
G^{*}_M]}{Re[ G_M G^{*}_E]} + \sqrt{\tau (\tau -1)} \frac{Re[G_M
\widetilde{F}_3]}{Re[ G_M G^{*}_E]} \cos \theta~, \nonumber\\
\delta(P_z) &=& \frac{2 Re[G_M \Delta G_M]}{|G_M|^2}- \sqrt{\tau
(\tau-1)} \frac{Re[G_M \widetilde{F}^{*}_3]}{|G_M|^2} \sin \theta
\tan \theta~.
\label{deltap}%
\end{eqnarray}
One can see both $\delta(P_x)$ and
$\delta(P_z)$ are the  odd functions of $\cos \theta$.

\section{Two-Photon Exchange Contribution}
This section is devoted to a directly numerical calculation for the
two-photon exchange. We know that much work has been done in the
space-like region. Naturally, it is expected that the same
calculation should be performed in the time-like region. After
considering the two-photon exchange, the amplitude $\mathcal{M}$
will be essentially modified, that is,
\begin{eqnarray}
\mathcal{M}=\mathcal{M}_0 +\mathcal{M}_{2\gamma},
\end{eqnarray}
where $\mathcal{M}_0$ is the contribution of the one-photon exchange and
$\mathcal{M}_{2\gamma}$ denotes the two-photon exchange. Therefore, to
the first order of $\alpha\ (\alpha = e^2/4\pi)$, we have,
\begin{eqnarray}
\frac{d \sigma}{d \Omega}\ \propto\ \overline{|\mathcal{M}|^2}\ =\
\overline{|\mathcal{M}_0|^2}\  (1+ \delta_{2 \gamma}) \nonumber
\end{eqnarray}
with
\begin{eqnarray}
\delta_{2 \gamma}=2 \frac{Re\{\overline{\mathcal{M}_{2\gamma}
\mathcal{M}_0^\dagger}\}}{|\mathcal{M}_0|^2}.
\label{delta}%
\end{eqnarray}
From the analysis in section 2, one can see that
$A_{2\gamma}(q^2,\theta)$ and $\delta_{2\gamma}$ are identical.

\par%

\begin{figure}
\centering
\mbox{\epsfig{figure=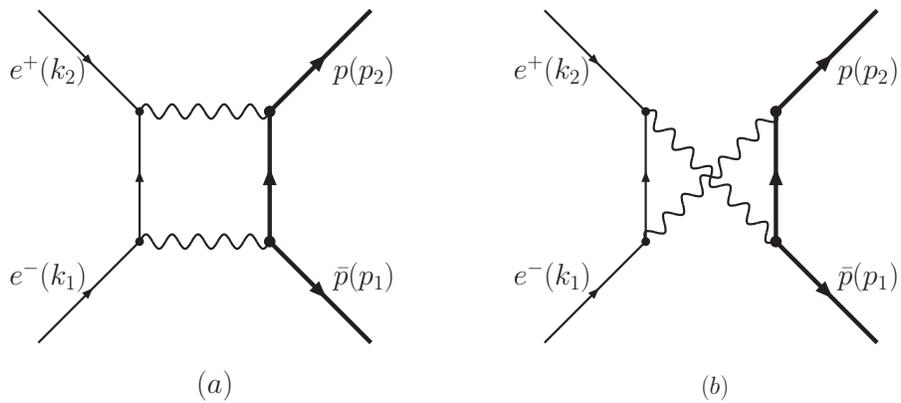,width=120mm,clip=}} %
\renewcommand{\figurename}{Fig.}
\caption{Two-photon exchange box and crossed box diagrams in
annihilating  channel.}
\label{Fig-feyntpe}%
\end{figure}

To proceed a  direct calculation, the amplitude of the two-photon
exchange from the direct box (Fig. \ref{Fig-feyntpe} $a$) and
crossed box diagram (Fig. \ref{Fig-feyntpe} $b$) has the form
\begin{eqnarray}
\mathcal{M}_{2 \gamma}= e^4 \int \frac{d^4 k}{(2 \pi)^4}\left[
\frac{N_{a}(k)}{D_a(k)}+\frac{N_{b}(k)}{D_{b}(k)}\right].
\label{mtpe}%
\end{eqnarray}
where the numerators are the matrix elements. For the direct box
diagram,
\begin{eqnarray}
N_{a}(k)= j_{(a)\mu\nu}J_{(a)}^{\mu\nu}~,\nonumber
\end{eqnarray}
with
\begin{eqnarray}
j_{a}^{\mu \nu}&=& \bar{u}(-k_2) \gamma^{\mu} (\hat{k}_1-\hat{k})
\gamma^{\nu} u(k_1),\nonumber\\
J_{a}^{\mu\nu}&=& \bar{u}(p_2) \Gamma^{\mu}(k_1+k_2-k)
(\hat{k}-\hat{p}_1-m_N) \Gamma^{\nu}(k) u(-p_1),
\end{eqnarray}
with $\hat{k} \equiv \gamma\cdot k$  and  $\Gamma_{\mu}(k)$ defined
in Eq. (\ref{ff0}). The denominators in Eq. (\ref{mtpe}) are the
products of the scalar propagators,
\begin{eqnarray}
D_{a}(k)&=&[k^2-\lambda^2][(k_1+k_2-k)^2-\lambda^2][(k_1-k)^2-m_e^2]
[(k-p_1)^2-m_N^2],
\end{eqnarray}
where an infinitesimal photon mass, $\lambda$, has been introduced
in the photon propagator to regulate the IR divergence. Similarly we
can write down the expressions of $N_{b}(k)$ and $D_{b}(k)$ for Fig.
\ref{Fig-feyntpe} $b$.

\par%

For the $1\gamma \otimes 2\gamma$ interference term, we define the
leptonic and hadronic tensors as,
\begin{eqnarray}
L^{(a,b)}_{\mu\nu\rho}=j^{(a,b)}_{\mu\nu} j^{*}_{\rho}~,~~~~
H^{(a,b)}_{\mu\nu\rho}=J^{(a,b)}_{\mu\nu} J^{*}_{\rho}\ .
\end{eqnarray}
Here the current operator in the hadronic current $J_{\rho}$ is the
same as the one in $J_{\mu\nu}$, then,
\begin{eqnarray}
\frac{d\sigma^{int}}{d\Omega}\propto\mathcal{M}_{2\gamma}\mathcal{M}_0
=\frac{e^6}{q^2} \int \frac{d^4k}{(2 \pi)^4} \Big[
\frac{L^{(a)}_{\mu\nu\rho} H^{(a) \mu\nu\rho}}{D_{a}(k)}
+\frac{L^{(b)}_{\mu\nu\rho} H^{(b) \mu\nu\rho}}{D_{b}(k)} \Big]
\label{mtpem0}.%
\end{eqnarray}
For the unpolarized differential cross section only
$L^{(a,b)}_{\mu\nu\rho}(0) H^{(a,b) \mu\nu\rho}(0)$ survives.  From
the crossing symmetry, we conclude that the expressions of
$\delta_{2 \gamma}$ are identical with Mandelstan variables for both
the scattering channel and the annihilating channel, that is,
\begin{eqnarray}
\delta_{2 \gamma}(s, t)_{e^{-} + p\rightarrow e^{-} +
p}=g(s,t)=\delta_{2 \gamma}(s, t)_{e^{-} + e^{+} \rightarrow  p +
\bar{p}}.
\end{eqnarray}
In the soft approximation $g(s,t)$ can be expressed as
\begin{eqnarray}
g(s,t)_{soft}=-2 \frac{\alpha}{\pi} \ln \left|
\frac{s-m_e^2-m_N^2}{s+t-m_e^2-m_N^2} \right| \ln
\left|\frac{t}{\lambda^2}\right|.
\label{soft}%
\end{eqnarray}
\par%

For the double spin polarization observables $P_x$ and $P_z$ the $1
\gamma \otimes 2\gamma$ interference contribution is
\begin{eqnarray}
P^{int}_{x,z}&=&\frac{e^2}{q^2 D} \int \frac{d^4k}{(2\pi)^4} \Big[
\frac{L^{(a)}_{\mu\nu\rho} H^{(a)\mu\nu\rho}(s_{1x,z})}{D_a(k)}
+\frac{L^{(b)}_{\mu\nu\rho} H^{(b)\mu\nu\rho} (s_{1x,z}) }{
D_b(k) }\Big]\nonumber\\
&=&\frac{e^2}{q^2 D} \int \frac{d^4k}{(2\pi)^4} \Big[
\frac{L^{(a)}_{\mu\nu\rho}(S) H^{(a)\mu\nu\rho}(s_{1x,z})}{D_a(k)}
+\frac{L^{(b)}_{\mu\nu\rho}(S) H^{(b)\mu\nu\rho} (s_{1x,z}) }{
D_b(k) }\Big].
\label{pint}%
\end{eqnarray}
As in the one-photon exchange approximation, the unpolarized
leptonic vector has no contribution to the double spin polarization
observables. For the term of $L_{\mu\nu\rho}(0) H^{\mu\nu\rho}
(s_{1x,z})$, after some algebraic calculations, we find that the
non-vanishing contributions are  in the forms of $\varepsilon^{a b c
k},\ a'\cdot k \varepsilon^{a b c k},\ a'\cdot k \ b'\cdot k
\varepsilon^{a b c k},\ k^2 \ a'\cdot k \varepsilon^{a b c k},$ and
$\{a',\ b',\ a,\ b,\ c\}\in \{s_{1},\ k_1,\ k_2,\ p_1,\ p_2,\}$.
Since $a',\ a,\ b,\ b' $ and $c$ have zero-$y$ component, then the
non-vanishing terms are the odd functions of $k_y$. Namely
$L_{\mu\nu\rho}(0)H^{\mu\nu\rho}(s_1) = f(s,t,k_0,k_x,k_z,k_y^2) k_y
$. Since the denominators in Eq. (\ref{pint}) are the even functions
of $k_y$, the contribution of $L_{\mu\nu\rho}(0)
H^{\mu\nu\rho}(s_1)$, therefore, vanishes.

\section{Numerical Results and Discussion}

In this work, we calculate the contributions of direct box diagram
(Fig. \ref{Fig-feyntpe} $a$ ) and crossed box diagram (Fig.
\ref{Fig-feyntpe} $b$ ) to the unpolarized differential cross
section and the double spin polarization observables. In this
calculation a simple monopole form of the form factors is employed.
This phenomenological form factor is $G_E(q^2)=G_M(q^2)/\mu_p=G(q^2)
= -\Lambda^2 / (q^2-\Lambda^2)$, with $\Lambda=0.84\ GeV$, which is
consistent with the size of the nucleon. Practically, for the
interaction of the outgoing hadrons, the time-like form factors have
a phase structure. which means the form factors are complex in the
time-like region. In this work what we concern is the ratio
$\delta_{2\gamma}$ and double spin polarization observables $P_x$
and $P_z$. Moreover, the phenomenological form factors appear in
both denominator and numerator of these physical observables. In
such cases, the form of form factors varies the ratio and
polarization observables in a very limited extension. The same
conclusion can be drawn from the results of two-photon exchange
corrections to space-like form factors in Ref. \cite{PRC-065203}.

\par%

In our calculation, the loop integrals of the two-photon exchange
contribution, firstly, were evaluated analytically in terms of the
four-point Passarino-Veltman functions \cite{NPB-151} using package
FeynCalc \cite{CPC-345}. Then, the Passarino-Veltman functions were
evaluated numerically with LoopTools \cite{CPC-153}. The IR
divergence in the $1 \gamma \otimes 2 \gamma$ is proportional to
$\ln \lambda$. This conclusion can be drawn by analyzing the
integral in Eq. (\ref{delta}) as well as by crossing symmetry and
previous results in the scattering channel. Furthermore, the
previous calculations in the scattering channel have shown that the
IR divergence in the two-photon exchange contribution is exactly
canceled by the corresponding terms in the bremsstrahlung cross
section involving the interference between the real photons emitted
from the electron and from the proton. With crossing symmetry, the
IR  divergence in the annihilating channel caused by the two-photon
exchange can also be ignored.

\par%

\begin{figure}[h]
\centering
\mbox{\epsfig{figure=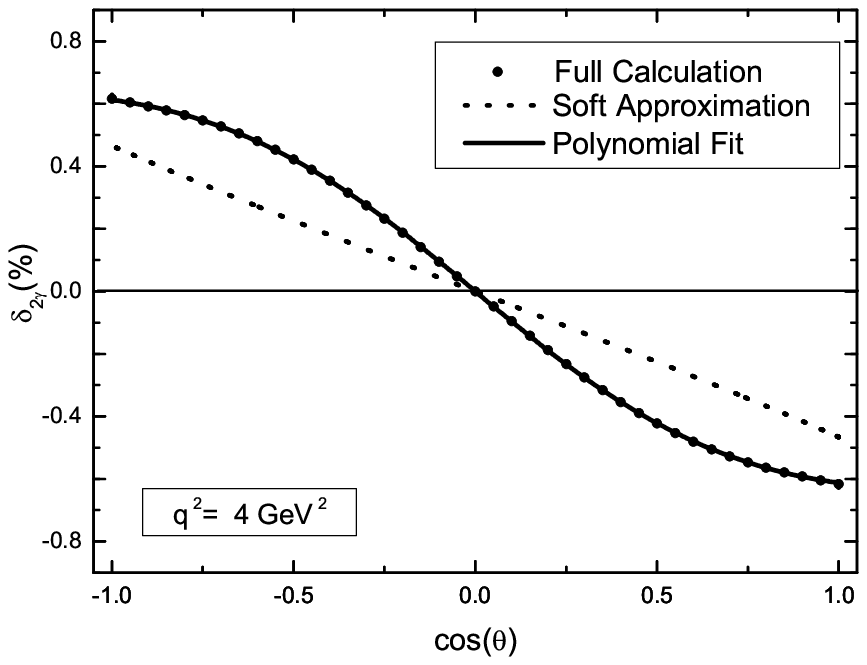,width=75mm,clip=}} %
\hspace{5mm}
\mbox{\epsfig{figure=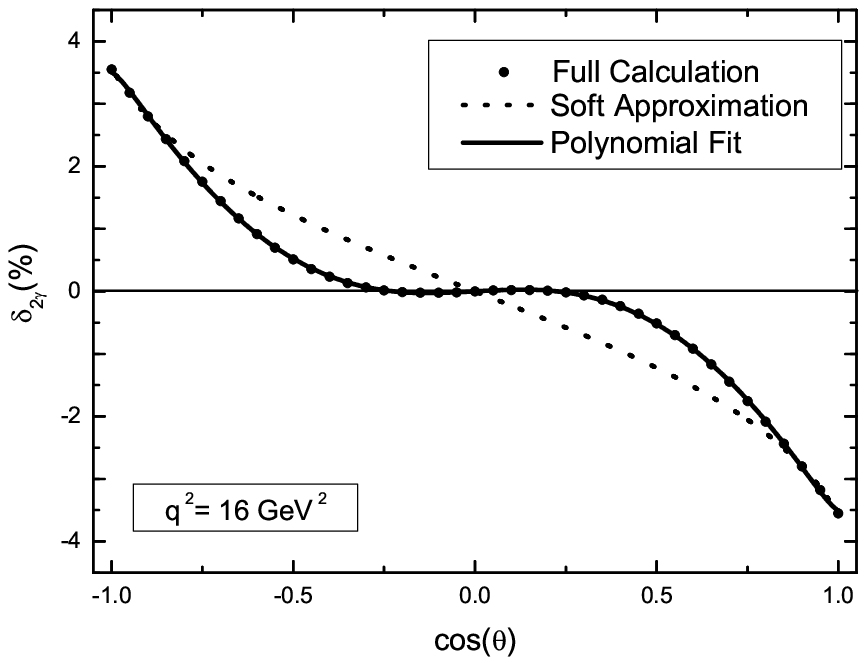,width=75mm,clip=}} %
\renewcommand{\figurename}{Fig.}
\caption{$\cos \theta$ dependences of the finite $2\gamma$
contribution to the unpolarized differential cross section. The full
circles are the results of full calculation, the dotted curves are
those with soft approximation, and the solid lines are the
polynomial fit for the full calculation. The left panel is the
result at $q^2=4~ GeV^2$ and right one is at $q^2=16~ GeV^2$.}
\label{Fig-unpolar}%
\end{figure}

From our previous analysis, the two-photon contribution to
unpolarized differential cross section $\delta_{2\gamma}$ is
identical to the angular asymmetry $A_{2 \gamma}$, which means
$\delta_{2 \gamma}$ is also the odd function of $\cos \theta$. The
numerical results of the two-photon contribution to unpolarized
differential cross section $\delta_{2 \gamma}$ are presented in Fig.
(\ref{Fig-unpolar}), where we show a comparison of $\delta_{2
\gamma}$ (defined as in Eq. (\ref{delta})) between the results of
the full calculation and the soft approximation. The full circles in
the figure are the full calculation, the dotted curves are the
results with soft approximation and the full curves are the
polynomial fits to the full calculation. We find a polynomial in the
form of $\cos \theta [a_0(t) + a_1(t) \cos^2 \theta + a_2(t) \cos^4
\theta + ...]$ can give a good fit with a power series of $\cos
\theta$ (no more than $\cos^5 \theta$). The left panel of Fig.
(\ref{Fig-unpolar}) shows the results with momentum transfer
$q^2=~4~GeV^2$, which is near the threshold of the reaction $e^{+}+
e^{-} \rightarrow p + \bar{p}$. We see that the two-photon exchange
contribution to the unpolarized differential cross section is rather
small, only about $\pm 0.6 \%$ at $\theta = \pi (0)$. In addition,
with the coefficients $a_0 = -9.6 \times 10^{-3}, a_1 = 4.9 \times
10^{-3}, a_2 = -1.5 \times 10^{-3}$ we see that the polynomial gives
a good fit of the full calculation. The right panel of Fig.
(\ref{Fig-unpolar}) is the results at $q^2= 16~GeV^2$, the
contribution of the two-photon exchange is relatively large, nearly
$4 \%$, the parameters of the fit for the full calculation are $a_0
= 2.9 \times 10^{-3}, a_1 = 5.7 \times 10^{-2}, a_2 = -1.9 \times
10^{-2}$. We conclude that at a fixed momentum transfer, the
contribution of the two-photon exchange is strongly dependent on
$\cos \theta$. In magnitude, the contribution is rather limited in
small momentum transfer region, with $q^2$ increasing, the
contribution becomes larger. This conclusion is consistent with the
results in the space-like region.

\par%

\begin{figure}[h]
\centering
\mbox{\epsfig{figure=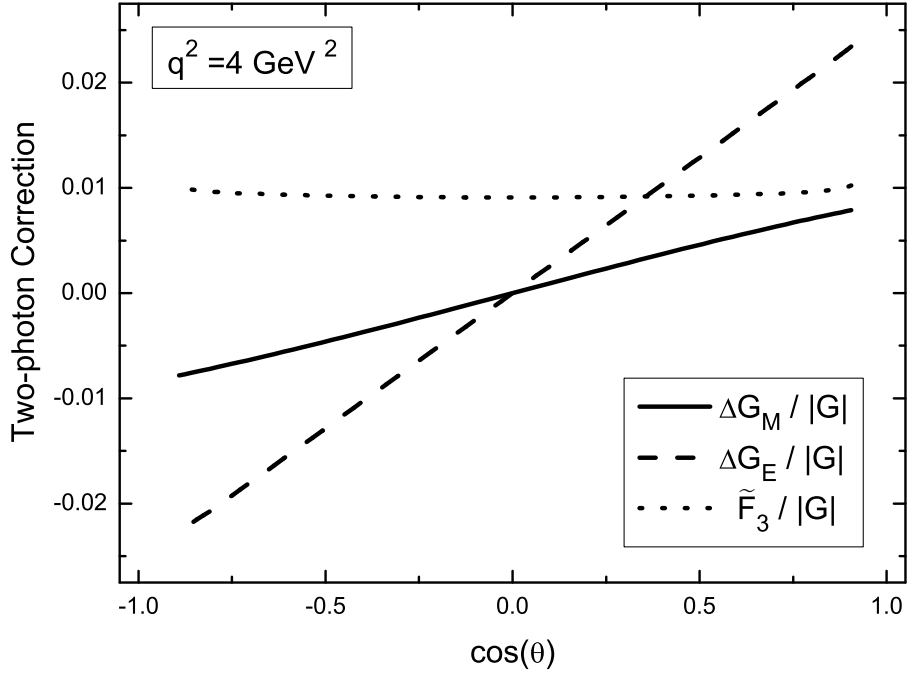,width=120mm,clip=}} %
\renewcommand{\figurename}{Fig.}
\caption{$\cos \theta$ dependence of the two-photon contribution to
the proton form factors in the time-like region at $q^2=4~GeV^2$.}
\label{Fig-ffs}%
\end{figure}

In Fig. \ref{Fig-ffs}, we show the $\cos \theta$ dependence of the
real part of corrections to the proton time-like form factors caused
by the two -photon exchange at $q^2~=~4 ~GeV^2$. For $\Delta G_E/G$
and $\Delta G_M/G$,  significant $\cos \theta$ dependences are
observed, while $\widetilde{F}_3/G$ weakly depends on $\cos \theta$.
For the parity, $\Delta G_E/G$ and $\Delta G_M/G$ are odd, and
$\widetilde{F}_3/G$ is even. These features are consistent with our
general analysis. The electric form factor is relatively more
sensitive to the two-photon exchange corrections, is about $2.5 \%$
at $\theta =0 (\cos \theta=1)$ and $-2.5 \%$ at $\theta = \pi (\cos
\theta=-1)$. The correction to magnetic form factor, $\Delta G_M /G$
varies from $1 \%$ to $-1 \%$ with $\theta$ from zero to $\pi$,
while $\widetilde{F}_3/G_E$ is about $1\%$ in the whole $\theta$
region.
\par%

\begin{figure}[h]
\centering
\mbox{\epsfig{figure=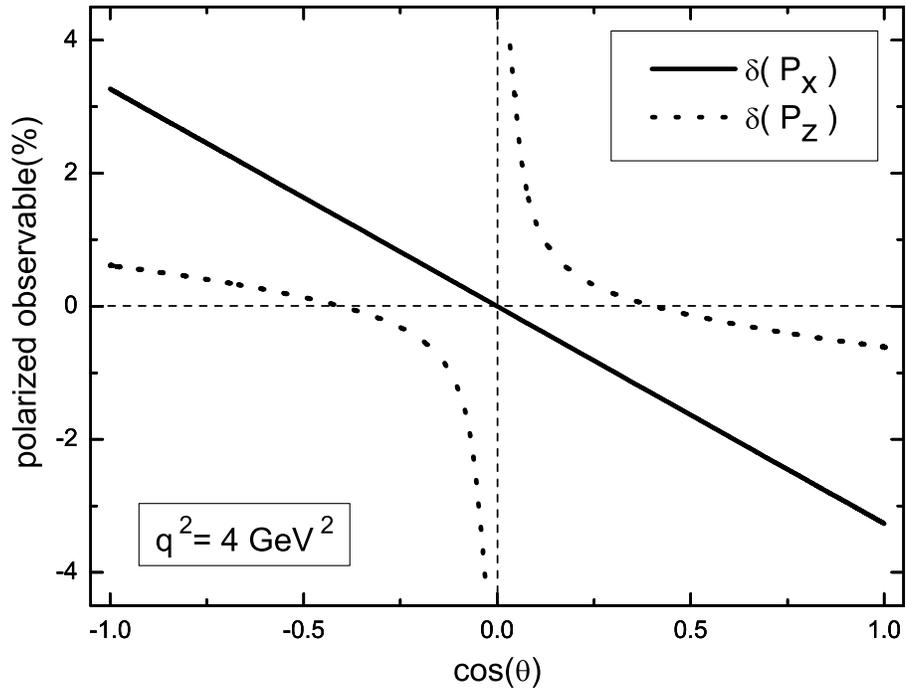,width=120mm,clip=}} %
\renewcommand{\figurename}{Fig.}
\caption{$\cos \theta$ dependences of the two-photon contribution to
the polarization observables at $q^2=4GeV$. The solid curve stands
for the results of $\delta(P_x)$, and the dotted curve represents
the results of $\delta(P_z)$.}
\label{Fig-polar}%
\end{figure}

From our previous analysis in Sec. \ref{dcspo}, one can see the
two-photon contributions to double spin polarization observables
$P_x$ and $P_z$ are even functions of $\cos \theta$. In our previous
numerical results in Fig. (4) we find $\widetilde{F}_3$ is not zero
at $\theta~=~\pi/2$, then $\delta(P_z)$ will be proportional to
$\tan \theta$ at the limit of $\theta \rightarrow \pi/2$ and will be
infinity at $\theta~=~\pi/2$. Our numerical results of the $\cos
\theta$ dependence of $\delta(P_{x,z})$ at $q^2~=~4~GeV^2$ are
displayed in Fig. (\ref{Fig-polar}). We can see that the two-photon
exchange contribution to the double spin polarization is strongly
$\theta-$ dependence, and is the odd function of $\cos \theta$,
which is consistent with our general analysis. For $P_x$, the
variation caused by the two-photon exchange reaches  maximum at
$\theta =\pi(0)$ (about $4 \%$). It seems that one can more easily
find the signal of the two-photon exchange at the backward ($\theta
= \pi$) and forward ($\theta = 0$) angle. However, notice Eq.
(\ref{polar}), we know that $P^{1\gamma}_x$ is proportional to $\sin
\theta$. It means when $\theta$ is very small(close to $0$) or very
large(close to $\pi$), $P^{1 \gamma}_x$ will be compatible to $0$,
and therefore, the absolute variation caused by the two-photon
exchange will be very limited. Thus, it will still be difficult to
find any signal of the two-photon exchange in the observable $P_x$.
For $P_z$, the contribution of the two-photon exchange reaches
maximum when $\theta= \pi/2$. In the one photon mechanism
$P^{1\gamma}_z $ is proportional to $ \cos\theta$, which suggests
that no matter what kinds of form factors we employed, $P^{1
\gamma}_z$ vanishes at $\theta ~=~\pi/2$. While taking the
two-photon exchange contribution into consideration, as in Eq.
(\ref{deltap}), $P_z(\pi/2)$ is not equal to zero any more. From the
experimental point of view, the nonzero $P_z$ at $\theta~=~\pi/2$
might be a strong evidence of the two-photon exchange in the process
of $e^{+} + e^{-} \rightarrow p + \bar{p}$.

\par%

According to our numerical results, one can see the two-photon
exchange contribution to the unpolarized differential cross section
$\delta_{2\gamma}$, which is identical to angular asymmetry
$A_{2\gamma}$, is rather small at small momentum transfer.  With
present experimental precision, it is rather difficult to find any
evidence of the two-photon exchange from the unpolarized observable
in the process $e^{+} + e^{-} \rightarrow p + \bar{p}$, especially
at low momentum transfer region. With $q^2$ increasing, the
contribution of the two-photon exchange becomes important. It can
be several percent at $q^2$ about $16 ~GeV^2$. Furthermore, for
the double spin polarization observables, $P_z$ deserves to be
considered in further experiment. In conclusion,the precise
measurements of the unpolarized differential cross section at high
momentum transfer and the double spin polarization observable $P_z$
especially at $\theta~=~\pi/2$ are expected to show some evidences
of the two-photon exchange in this process.

\par%

\section{Acknowledgment}
This work is supported by the National Sciences Foundations of China
under Grant No. 10475088, No. 10747118, and by CAS Knowledge
Innovation Project No. KC2-SW-N02.

%%%%%%%%%%%%%%%              Reference             %%%%%%%%%%%%%%%%%%%%%

%%%%%%%%%%%%%%%               Figures              %%%%%%%%%%%%%%%%%%%%%
\clearpage\newpage%
%%%%%%%%%%    Fig. 1  %%%%%%%%%%%%
%%%%%%%%%%    Fig. 2  %%%%%%%%%%%%

%%%%%%%%%%    Fig. 3  %%%%%%%%%%%%
%%%%%%%%%%    Fig. 4  %%%%%%%%%%%%
%%%%%%%%%%%    Fig. 5  %%%%%%%%%%%%

\end{document}